\title{Muon excess at sea level from solar flares in association with the Fermi GBM spacecraft detector}
\author{
        C.R.A.Augusto, C.E.Navia, K.H.Tsui, H.Shigueoka \\
                Institito de Fisica
        Universidade Federal Fluminense\\
        24210-336, Niteroi RJ, \underline{Brazil}
            \and
        A.C.Fauth\\
        Instituto de Fisica Gleb Wataghin\\
        Universidade Estadual de Campinas
        13083-970, Campinas SP, \underline{Brazil}
}

\documentclass[12pt]{article}

\usepackage{graphicx}

\begin{document}
\maketitle

\begin{abstract}
This paper presents results of an ongoing survey on the associations between muon excesses at ground level registered by the Tupi telescopes and transient solar events, two solar flares whose gamma-ray and X-ray emissions were reported by, respectively, the Fermi GBM and the GOES 14. We show that solar flares of small scale, those with prompt X-ray emission classified by GOES as C-Class (power $10^{-6}$ to $10^{-5}$ 
W m$^2$ at 1 AU) may give rise to muon excess probably associated with solar protons and ions emitted  by the flare  and arriving at the Earth  as a coherent particle pulse. 
The Tupi telescopes are within the central region of the South 
Atlantic Anomaly (SAA), which allows particle detectors to achieve a low rigidity of response to primary and secondary charged particles ($\geq 0.1$ GV). Here we argue for the possibility of a ``scale-free'' energy distribution of particles accelerated by solar flares. 
Large and small scale flares have the same energy spectrum up to energies exceeding the pion production, the difference between them is only the intensity.
If this hypothesis is correct, the Tupi telescope is registering muons produced by protons (ions) whose energy corresponds to the tail of the spectrum. Consequently the energy distribution of the emitted protons has to be a power law spectrum, since power law distributions are characterized as scale free distributions. The Tupi events give support to this conjecture  \ldots
\end{abstract}

\section{Introduction}

The last solar cycle 24 started at the beginning of 2008 \cite{svalgaard09}. Even so,
in this second semester of 2010, we are still at a low level of activity,
 in an anomalous extended period of minimal solar activity.
But sunspots are starting to appear again, and spots are the manifestation of 
the magnetic field poking through the surface of the Sun.

This Sun cycle anomaly is the first in the spatial era, i.e., where the Sun is monitored by 
spacecraft detectors, but there are such
registered from 107 years ago, of a similar pattern. That happened during the transition 
between cycles 13 and 14.

Solar flares are among the most powerful astrophysical phenomena in the 
solar system.
Initially the observation and the study of these solar flares 
used detectors on the surface of the Earth, mainly by neutron monitor experiments. A lot  has 
been obtained with these observations, such as the anti-correlation between solar activity 
and the flow of galactic cosmic rays. The existence of a prompt and late emission in flares and their 
correlations with Coronal Mass Emission (CME), Forbuch events, a fall in the cosmic ray intensity, due to
a solar disturbance crossing the Earth, and so forth.
.

However, in most cases, the observations are restricted to flares with high intensity,   
those with an X-ray flux at 1 UA is classified by GOES as X-Class and M-Class flares, 
flux above $10^{-4}$ and $10^{-5}$ W m$^2$ respectively. 
Recently the Tupi experiment has reported   
experimental evidence of muon excess in association with high energy particles (protons and ions) with energies above   
the pion production threshold (because they produce muons in the atmosphere), emitted by flares  
of small scale, those with an X-ray flux below  $10^{-5}$\ W m$^2$ or C-Class flares \cite{navia05,augusto05}.

Even in events observed at ground level in association with large solar flares, the acceleration mechanism producing
particles (protons, ions) up to several tens of GeV is not well understood. The situation becomes still
more critical in the case of ground events associated with solar flares of small scale. Here we argue for the
possibility of a ``scale-free'' energy distribution of particles accelerated by solar flares.
This mean that high and low fluxes of solar particles, associated to big and small flares, have the 
same energy spectrum, up to energies exceeding the pion production. The difference between them is only the intensity.

This hypothesis is corroborated by new observations of two solar flares of small scale observed by 
Tupi telescopes, in association with energetic gamma-rays 
detected by Fermi GBM (designed to observe gamma-ray bursts) and with X-ray flux detected by GOES 14;
both are spacecraft detectors orbiting the Earth.

\paragraph{Outline}
This paper is organized as follows:
In Section 2 a brief description of the Tupi experiment is presented
and includes a comment on the location of the Tupi telescopes.
 In Section 3, we argue why the telescopes have a high sensitivity.
Section 4 contains a brief description of micro and mini solar flares.
A brief study on muon excess and solar flare association is presented in Section 5
Section 6 is devoted to showing the results of the association
 between muon excesses at ground level and two solar mini flares,
 registered by GOES-14 (X-ray flux) and Fermi GBM (gamma-ray counting rate). 
 Section 7 contains an analysis of the pitch angle and muon excess intensity,
 and Section 8 contains conclusions and remarks.

\section{The Tupi experiment}
\label{sec2}

The Tupi experiment \cite{augusto08} is an Earth-based muon apparatus,
 devoted to the study of cosmic rays,
 located in Niteroi City, Rio de Janeiro, Brazil (22.88S, 43.16W), consequently
 near the SAA central region.
 The apparatus has two telescopes.
 Each telescope was constructed on the basis of two detectors, plastic scintillators, 
 0.5 m$\times$ 0.5  $\times$ 35 cm, placed perpendicularly to the axis of the telescope. The distance 
 between the plastic scintillators is 3.0 m.
 
 The general layout of the vertical telescope,
 including the data acquisition, is shown in Fig.1. The data acquisition system is made on the 
 basis of a USB card, with a counting rate of up to 100 kHz per channel.  All steps from signal 
 discrimination to the coincidence and anti-coincidence are made via software, using the virtual 
 instrument technique. The application programs were written using the Lab-View tools.

 The main task of the first level trigger is a coincidence
 between the two detectors. The second-level trigger is a veto for air shower coming from 
 other directions, and contains a third detector placed off the telescope axis. Thus the directionality 
 of each muon telescope is guaranteed by a veto or anti-coincidence guard,
using a third detector. Therefore, only muons with trajectories close to the telescope axis are registered.

One of the two telescopes has a vertical orientation,
 and the other is oriented near 45 degrees to the vertical (zenith)
 pointing to the west.
 Both telescopes have an effective aperture of 65.6 cm$^2\;$sr,
 projecting to the space a cone with an open angle of 9.5$^0$ in relation of the telescope axis.

At sea level and in the sub-GeV to GeV energy region, the muon flux is $\sim 70$ times higher than 
the nucleon flux, and $\sim 800$ times higher than the electron flux. Consequently,
the telescopes are capable of detecting  mainly muons induced in the atmosphere by cosmic rays 
(mostly protons). The minimum proton energy $E_{pth}$ needed to produce muons of energy $E_{\mu}$ in the atmosphere is $E_{p th} \sim 10 \times E_{\mu}$. In the case of the Tupi experiment, the muon energy threshold is $E_{\mu} \sim $0.1 GeV . This means that the minimum proton 
energy is $E_{pth} \sim $1.0 GeV.

The telescopes are inside a building and there are two flagstones on the telescopes and only muons
with an energy higher than 0.1 GeV are detected. This muon energy threshold is that required to 
penetrate the two flagstones. The flagstone reduces the noise due to other non-muon particles, for
example, it is opaque to electrons.

The Tupi experiment has a fully independent power supply,
 with an autonomy of up to six hours
 to safeguard against local power failures.
 As a result, the data acquisition is carried out 24 hours a day,
 giving a duty cycle greater than 90\%.

\subsection{Location of the Tupi telescopes}
\label{sec21}

The Earth is surrounded by the magnetosphere
 which protects us from cosmic rays with energies less than several GeV
 by deflecting or capturing them in the Van Allen radiation belts.
 Even so, there is an additional factor
 that may locally enhance the cosmic ray intensity at middle latitudes.
 This is the so-called South Atlantic Anomaly (SAA),
 which is an area of anomalously weak geomagnetic field strength.
 
 In the IGRF95 data \cite{barton97}, the magnetic field strength in the central SAA region 
(26S, 53W) is 24,000 nT. This area coincides with the Atlantic coast to the south-west of Brazil,
while the total SAA area covers a great part of South America's central region. 
 The boundary region is  defined as that  
 where the magnetic field strength is less than 28,000 nT, as shown in Fig.2 (top panel).  
 
  The SAA is a result of the eccentric displacement
 of the magnetic field centre from the geographical centre of the Earth (by about 400 km)
 as well as the displacement between the magnetic and geographic poles of the Earth.
 This behaviour permits the inner Van Allen belt
 to impart highly energetic particles (mostly protons),
 penetrating deeper into the atmosphere
 owing to the low field intensity over the SAA,
 and thereby interacting with the dense lower atmosphere,
 resulting in higher ionization and increased electrical conductivity \cite{abdu05}.
 
 Thus, cosmic rays with an energy around several dozens of MeV to GeV
 including solar particles
 can reach the north and south polar regions,
 but it is difficult for them to reach low latitude regions
 due to a high rigidity cut-off (above 9 GV).
 Nevertheless, in the SAA area, the anomalously weak geomagnetic field strength
 gives the muon telescopes the lowest rigidity of response
 to cosmic protons and ions, ($\geq 0.1$ GV).
 
According to satellite data, such as from the MOPPIT space instrument \cite{nichitiu04}, the geographical distribution of single events (particles reaching the Earth) presents a great anisotropy with a much higher event density in the SAA region. The background events not connected to the SAA or Polar regions (20\%) are due to galactic cosmic rays. Single events due to particle precipitation contribute 54\% in the SAA (Van Allen background) and 26\% in the Polar regions. Most of the single particle events observed by satellite instruments are in the keV to MeV energy range. The MOPPIT result of the geographical distribution of single events is summarized in Fig.2 (bottom panel).

It has been shown that the SAA region \cite{augusto10} also favours the precipitation of high energy particles (Van Allen background), with energies above the pion production threshold, because they produce air showers in the Earth's atmosphere and the hard component (muons) is able to reach sea level \cite{augusto10}. 
Even so, the Van Allen background has an enhancement only at daytime, beginning  three hours after sunrise until around one hour after sunset. 
Because the focus effect of the interplanetary magnetic field (IMF) lines become dominant, in this schedule, they point for the Earth surface.


\section{Particle telescope sensitivity}
\label{sec3}

The Earth's magnetic field deflects the charged particles
 of the shower initiated by a primary particle.
 This deflection is caused by the perpendicular component
 of the Earth's magnetic field to the particle trajectory.
 This effect results in a decrease in the number of collected particles
 and therefore in telescope sensitivity.
 This means that the sensibility of particle telescopes
depends of the transverse component of the  magnetic field intensity of the Earth.
 
The radius of curvature, $R$,
 of a positive muon travelling down the atmosphere with momentum $p$
 perpendicular to the Earth's transverse magnetic field, $B_{\bot}$,
 is $R=p/(eB_{\bot})$.
 As the muon travels, it will be shifted horizontally
 by a distance $\delta x$ in the direction perpendicular to $B_{\bot}$.
 A formula for $\delta x$, to first order in $z/R$
 where $z$ is the height of the atmosphere where the muon is generated,
 can be obtained as $\delta x \sim z^2/R= z^2ceB_{\bot}/p$.
 Thus the muons are shifted by a quantity
 that depends on the distance $z$
 but also on the momentum $p$ of the muon.
 The $\Delta x$ in the SAA area is at least 50\% smaller
 than the $\Delta x$ outside the SAA area.
 Fig.3 summarizes the situation.
 This means that the sensitivity of the telescopes
it is up to three times higher in the SAA region
 because the transverse magnetic field is very small,
 even smaller than the average value for the polar regions.
 This offers the Tupi muon telescopes
the opportunity to observe small scale transient events.

 On the other hand, in the Earth's upper atmosphere, there is a layer called the ionosphere (100 km above sea level) which is characterized by a larger ion concentration. According to Chapman's hypothesis \cite{mitra38} regarding the different origins of the ionosphere, the F ionospheric layer is attributed to the action of the ultra-violet rays of the sun and the E layer is attributed to the action of neutral solar corpuscles. However, in the stratosphere at low and middle latitudes and starting from 12 km above sea level there is a residual ion concentration \cite{gupta04}. The ionization in this region is mainly due to cosmic rays and the effect is an increase in the electric conductivity of the stratosphere, which is given by
\begin{equation}
\sigma=-\frac{e^2}{\nu}\sum\left( \frac{n_i}{m_i}\right),\label{eq2}
\end{equation}
where $e$ is the electron charge, $\nu$ is the collision frequency, $n_i$ is the ion density and $m_i$ is the ion mass.

In the SAA region, due to the influence of the lower Van Allen belt, which imparts highly energetic particles (mostly protons) coming deeper down into the atmosphere, there is an enhancement of the ionization production and increasing electric conductivity, and it embraces great atmospheric depths. A consequence of this behaviour is the change in the value of the coefficient that describes the energy loss of a charged particle such as a muon for ionization. The result is an increase in the range of charged particles.
For instance,  in the SAA region, cosmic ray fluxes are even higher than world averages at comparable altitudes,
reflecting an enhancement of incoming primary cosmic rays.

 \section{Solar flares of small scale}
 \label{sec4}
 
A solar flare is defined as a sudden, rapid, and intense variation in brightness.
 A solar flare occurs when magnetic energy
 that has built up in the solar atmosphere
 is suddenly released.
 Radiation is emitted across virtually the entire electromagnetic spectrum.
 In solar flares, the interaction of the energetic electrons
 with thermal protons provides the deceleration,
 and X-ray photons with energies less than
 or nearly equal to the electron energy are produced.
 These X-ray photons are the emitted radiation signatures
 detected by scientific instruments such as GOES and SOHO.
 The frequency of flares coincides with the eleven year solar cycle.
 When the solar cycle is at a minimum,
 active regions are small and rare and few solar flares are detected.
 Their occurrence increases in number
 as the Sun approaches the maximum of its cycle.
 However, the period around the solar minimum
 is useful for the observation of small transient events,
 such as micro and mini flares, whose flux is less than $10^{-5}\;$Watts m$^{-2}$.

Harder X rays with energies greater than 10 keV are also believed 
to be electron-ion bremsstrahlung. Spectral measurements of such hard X rays follow a
power-law rather than an exponential shape.

However, the gamma radiation observed in a solar flare has several origins.
The bremsstrahlung spectrum can extend up into 
the gamma-ray range. Indeed, in some of the biggest flares, the spectrum is seen
 to extend to energies in excess of 100 MeV. Proton and heavy ion interactions
also produce gamma-rays through $\pi^0$
decay, resulting in a spectrum that has a maximum at 68 MeV.

At higher energies, there are gamma-rays produced, not from the flare electrons,
 but from nuclear interactions of the protons and heavier ions accelerated in the flare,
or in other words, released by a flare driven shocks.
These high energy particles interact with the nuclei of the different elements in the
ambient solar atmosphere to produce a far more complicated emission spectrum. A continuous
spectrum and many individual gamma-ray lines have been identified, and they result from the decay of
elements in the solar atmosphere as carbon, nitrogen, etc.,
that are excited to high energy states in various nuclear interactions.

 \section{Association between muon excess and solar flare}
 \label{sec5}
 
 Energetic particles from solar flares,
 moving along solar magnetic field-lines,
 undergo pitch angle scattering
 caused by magnetic field fluctuations.
 This process is often assumed to be the basic physical process
 behind diffusive propagation of solar particles
 in interplanetary space,
 and they are also subjected to adiabatic losses.
 Fig.4 summarizes the situation,
 where the diffusive transport between the Sun and the Earth
 of solar charged particles emitted in a flare is shown. They are the 
 two equi-density curves, they open up (they diffuse)  when they spread sun away.
 Consequently, in most cases, only particles produced in very intense flares
 will give some signal (bigger than the background produced by the cosmic rays)
 in detectors located on the Earth's surface.
 
 The association between an X-ray flare and a muon excess requires taking into account the delay
 between the flare start and the time of flight between the Sun and the Earth of energetic particles.
 The time of flight is estimated from results of simulation under the assumption of which the propagation
 of energetic particles released by a flare driven shocks are injected  to the planetary medium in coherent pulses
 of energetic particles and a realist Archimedean spiral field lines around the Sun \cite{ruffolo95}. Here,
 we summarize simulation results for a typical arc length along the magnetic field (garden hose direction).
 $<z>(=1.3\;$AU), as a function of the distance travelled, $S$, which can be expressed as
\begin{equation}
 <z>=\alpha (\lambda)S.
 \end{equation}
 
 The constant $\alpha(\lambda)$, depending on the scattering mean free path, $\lambda$, and 
 satisfying the constraint condition
 $\alpha (\lambda)=1$ for $\lambda=1.0$\ AU, sometimes called the `scatter-free' condition, implies 
 that particles freely stream along the field at their maximum speed like a coherent particle pulse.
 In other words, the focus effect of the interplanetary magnetic field (IMF) lines become dominant and 
 the propagation of the energetic particles is like coherent pulses, following trajectories around the 
 field lines of the IMF.
 
 In contrast, in the other extreme case, when the scattering mean free path is small compared to the
 scale length of the IMF (i.e., $\lambda \leq 0.2$\ AU), particle propagation follows helical 
 trajectories around the IMF. Fluctuations of small scale in the IMF act as scattering centres of 
 the particles and the propagation is dominantly diffusive.
 
 A muon excess, detected at sea level by a directional telescope, is the signature of primary particles
 arriving at the top of the atmosphere with a strong anisotropic pitch angle distribution: `an almost
 coherent pulse'.  The rise time in the profile time of the muon excess can be used to infer
 the coherency of a pulse: for instance a very short rise time suggests a non-diffusive coherent 
 particle pulse transport. Thus an almost coherent pulse has a  value of $\lambda \geq $0.2 UA and 
 corresponds to a time of flight of energetic particles with a mean rigidity of 1 GV of up to 35 minutes.
 This delay correspond to up to $\sim 27$ minutes in relation to the X-ray signal at 1AU.
 
 Ground-level solar flares are usually observed
 by high latitude neutron monitors at relatively low rigidities ($\sim$ 1--3 GV),
 and in most cases the ground-level events are linked to solar flares of high intensity
 whose prompt X-ray emission is catalogued as X-class (above $10^{-4}$\ Wm$^{-2}$).
 Evidently, solar flare detection at ground level depends on several aspects,
 such as a good magnetic connection between the Sun and Earth.
 Most solar flares associated with GLEs
 are located on the western sector of the Sun
 where the IMF is well connected to the Earth.
 An example of a good magnetic field connection
 between the Sun and Earth is shown in Fig.5.
 Flares located near the foot-point of the ``garden hose'' field line
 between the Sun and Earth
 reach the Earth with a pitch angle close to $45^0$,
 because protons (ions) travel toward the Earth in a spiral trajectory,
 following the garden hose field line.
 
\section{Results}
\label{sec6}

The GOES14 Ion Chamber Detectors provide whole-sun X-ray fluxes for 
the 0.5--4.0 and 1.0--8.0 $\dot{A}$ wavelength bands.  In most cases the GOES
X-ray fluxes of solar flare are used as a reference for other types of detection,
such as ground level detectors and the Fermi GBM.

In addition, the Fermi GBM detector is designed to observe
gamma-ray bursts (GRBs) in the FOV (field of view)
of the GBM instrument, that is, if the GRB is bright enough to be localized, and in the energy 
range 10 keV to 30 MeV.

Here we will begin analysing the time profiles of the muon counting rate of these two muon excesses 
observed on 4 October 2010 and 3 November 2010 and 
compare them with the time profiles
of, respectively, the GOES14 X-ray fluxes and Fermi GBM gamma-ray counting rate.

\subsection{The event on 4 October 2010}
\label{sec61}
 
  The muon enhancement on 4 October 2010 is
 a sharp peak in the muon counting rate ($E_{\mu}>$0.1 GeV)
 observed in the vertical muon telescope. The onset in time 
 correlation has great statistical significance ($7.1\sigma$) when the counting rate is
 5 minutes binning, the muon arrival excess is $12.42$ minutes late in relationship 
 with the arrival of an X-ray excess flux  on GOES14. This is a
 mini flare whose X-ray prompt emission is classified as 1.9C-class
 (a flux of $1.9\times 10^{-6}$\ Watt/cm$^2$).
 Fig.6 summarizes the situation,
 where the GOES14 X-ray prompt emission on 4 October 2010
 for two wave lengths is shown in the top panel,
 and the corresponding (vertical) Tupi muon counting rate
 (5 minutes binning) is shown in the bottom panel. There is no signal in 
 the inclined telescope.
 
 The time delay between the X-ray detection
 and muon excess detection is 12 minutes.
 This means that the time of flight from the Sun to the Earth
 of the energetic particles was 20 minutes. According to the criteria 
 established in the last section, the muon excess was formed by primary particles
 arriving at the top of the atmosphere with an anisotropic pitch angle distribution, `an almost
 coherent pulse', and with a rigidity above 1 GV (1 GV corresponds to 1 GeV for protons).
 
 On the other hand, solar flare events have been searched for since 2008 by the Fermi GBM.  
So far, even in this minimum solar activity period, the Fermi GBM has found evidence of a high energy  photon emissions (gamma-rays)  
from solar flares. From 27 October 2009 to 3 November 2010, 22 solar flares have been registered by the Fermi  GBM. Solar activity is expected to rise in the next months, reaching a maximum in 2012, and the number of flares observed by the GBM must increment by a factor of up to ten for the same period.

For the first flare here analyzed and observed on 4 October 2010, the Fermi GBM trigger was at 16:32:31.443 and it was classified as a solar flare with a reliability of 0.8. The Fermi GBM light curve is shown in Fig.7 (top panel), it has a very fast rise time, even a faster rise time than the emission of X-rays observed by GOES 14 (see Fig.6, top panel). In contrast the rise time observed in the time profiles of the muon counting rate is not so fast as is shown in Fig.7 (bottom panel), this shows a certain degree of diffusion of the energy particles emitted by the flare during its transport. 

However, the flare duration (gamma-ray emission) in the Fermi GBM is very short when compared with the X-ray emission
and the muon counting rate. The gamma-ray emission as observed by Fermi does not exceed 467 seconds, while the X-ray emission and
the muon excess  has a duration bigger than 2000 seconds. In addition,
it is possible to infer up to three peaks (or at least two with high confidence) in the Fermi GBM light curve.   
These peaks seem to have a correlation with the three peaks observed in the muon counting rate. Even so, they have a temporal shift.

\subsection{The event on 3 November 2010}
\label{sec62}

The muon enhancement on 3 November 2010 has
 two sharp peaks in the muon counting rate 
 observed in the vertical muon telescope. The onset in time correlation 
 has great statistical significance of up to $20.0\sigma$ when the counting rate is
 5 minutes binning, the arrival of the muon excess is also 12 minutes late in relationship 
 with the arrival of an X-ray excess flux  at GOES14 as is shown in Fig.8. This mean that the muon excess 
 was formed by primary particles
 arriving at the top of the atmosphere as a
 coherent pulse, and with a rigidity above 1 GV. We see again that no signal exists in the inclined telescope.

 This flare is more intense than the previous one, it is of 5.0C-class. In addition, it is possible 
 to see the same structure in both the time profiles (X-ray flux and muon counting rate): two serial peaks. 
The time interval between these two  serial peaks in both GOES 14 X-ray flux and Tupi muon excess is $\sim$ 38 minutes.
That is a signature of there being little diffusion of the particles during their transport from the Sun to the Earth following IMF lines.

For this second flare, observed on 3 November 2010, the Fermi GBM trigger was at 12:13:10.916 and it was classified as a solar flare with a reliability of 0.92. The Fermi GBM light curve is shown in Fig.9 (top panel) and in the same figure are (bottom panel) the time profiles (raw data) of the muon counting rate. In this case both the muon excess and the GOES 14 X-ray flux have a faster rise time than the rise time observed in the Fermi GBM light curve. The gamma-ray emission duration as observed by Fermi GBM is short (1295 seconds) in relation with the muon excess duration of around 1 hour. This contrasts with the long duration of the X-ray emission as observed by GOES 14, because after a fast rise time, it falls off slowly. However the two peaks observed in the muon counting rate, during the first 500 seconds as is shown in Fig.9, are in excellent correlation with the two peaks observed in the Fermi GBM. 

In addition, the time delay of the muon excess in relation with the gamma-ray detection by Fermi (at 1 AU) is less than 33 seconds. Again, this is a signature indicating  a non-diffuse transport of the charged particles emitted by the solar flare.

\section{Pitch angle and muon excess intensity}
\label{sec7}

Fig.10 shows the Sun location, at Fermi trigger time, and the muon excess location during the sky scanner by the vertical telescope due to the rotation of the Earth, for the case of the two flares here analysed. The size (diameter) of the open circles is proportional to the logarithm of the muon counting rate. In both, the difference between the Sun and telescope axis declination is small, $\Delta \delta = 17.2 $ and  $\Delta \delta = 7.5 $,
respectively. Thus the pitch angle defined as the Sun-ward direction and the telescope axis is approximately equal to the difference between the right ascension of the Sun and of the axis of the telescope. Here a zero degree pitch angle represents the telescope pointing in the Sun-ward direction.
We would like to point out that one hour in right ascension corresponds to 15 degrees.

Following Fig.10, we can observe a pitch angle around 18 degrees for the first event and 51 degrees for the second event.  Flares located near the foot-point of the `garden hose' field line between the Sun and Earth
 reach the Earth with a pitch angle close to $45^0$, as is shown in Fig. 5 (that is, a good magnetic connection between the Sun and Earth).
Consequently, the charged particle transport is along an IMF line: under this condition, the particles will be subject to only small fluctuations
 of the IMF and the transport will be practically non-diffuse.
 
 This means that the diffusion of the particles carried during their transport between the Sun and the Earth  
is smaller for the second event. In fact, the rise time in the time profiles of the muon counting rate is smaller for the second event.  
In addition, for the second event the interval among the peaks observed basically is the same of that observed in GOES 14 and Fermi GBM.
Consequently, a coherent pulse of particles arriving in the top of the atmosphere was responsible for the excess of muons observed in the vertical telescope. 

On the other hand, a more prolonged rise time, as observed in the time profiles of the muon excess of the first event, means a more accentuated diffusion of the particles: this happens because the particles were ejected in space by the Sun off of the direction of an IMF line crossing the Earth, and they have a pitch angle around 18 degrees.  
Even so, a good correlation exists among the excess of muons observed in the vertical telescope and the emission of X-rays  observed by the GOES 14 and the emission of gamma-rays as observed by the Fermi GBM.  

\section{Conclusions and remarks}
\label{sec8}

Solar flares release energy in many forms---electro-magnetic (gamma-rays and X-rays), 
particles (protons, ions and electrons), and mass flows. Flares are characterized by their 
brightness in X-ray flux. The biggest flares are X-Class flares (flux between 10$^{-4}$ and   10$^{-3}$\ Watt m$^2$). M-Class flares 
have one tenth the energy and C-Class flares have one tenth of the X-ray flux seen in M-Class flares. 

In most cases only flares of X-class and M-class are observed by ground detectors. However, we report here two C-class flares
observed as muon excesses by the vertical Tupi telescope located at sea level and within the South Atlantic Anomaly (SAA) region.
We argue here that the high sensitivity attained by the Tupi telescopes is a consequence of its location, as in the SAA region
the shielding effect of the magnetosphere is not perfect and shows a `dip'. The SAA is an area of anomalously weak geomagnetic 
field strength.
This characteristic offers muon telescopes (inside the SAA region) the possibility of achieving a low rigidity of response to 
primary and secondary charged particles ($\geq 0.1$  GV).

We have shown that both events (muon excess) are in excellent correlation with the X-ray emission observed by GOES 14, as well as
in excellent correlation with the gamma-ray emission observed by Fermi GBM. In addition, from an analysis on the basis of a Monte Carlo
study, rise time in the  muon time profiles, and their pitch angle, we conclude that the second event on 03 November 2010 is constituted
 at least by two extremely coherent muon pulses, which means that the particles producing muons in the Earth atmosphere were emitted by the 
 Sun in the same direction as an IMF line crossing the Earth and their transport was practically non-diffuse.

Already for the first event analysed, and registered on 4 November 2010, the transport conditions between the Sun and the Earth were not very favourable. Even so it has a good correlation with both the Goes 14 X-ray flux and the Fermi GBM gamma-ray counting rate.

If muons are detected at sea level with energies above 0.1 GeV, this means primary (protons) with energies above 
the threshold of pion production ($\sim 1 GeV$). Even so, what is the mechanism in the flare that accelerates protons to these energies? Especially for flares of small scale, such as those of C-Class, is this a question.

Here we have argued for the possibility of a `scale-free' energy distribution of particles accelerated by solar flares.
Large and small scale flares have the same energy spectrum, up to energies exceeding the pion production,
 they differ only in their intensity.

If this hypothesis is correct, the vertical telescope is registering muons produced by protons (ions) whose energy corresponds to the tail of the spectrum. Consequently, the energy distribution of the emitted protons has to be a power law spectrum, as power law distributions are characterized as scale free distributions.

Finally, we would like to point out that solar flare events also have been searched for in the first
year of  Fermi LAT data (August 2008-August 2009). Up
until now there has been no evidence of high energy emission
from solar flares detected by the LAT \cite{iafrate09}.

\section{Acknowledments}

This work is supported by the National Council for Research (CNPq) of Brazil, under Grant  $479813/2004-3$ and $476498/2007-4$.
We are grateful to the various catalogs available on the web and to their open data policy, especially to the Space Weather 
Prediction Center (SWPC) and the Fermi GBM catalog.

\bibliography{simple}

\newpage

\begin{figure}
\vspace*{-6.5cm}
\hspace*{-2.0cm}
\centering
\includegraphics[width=18.9cm]{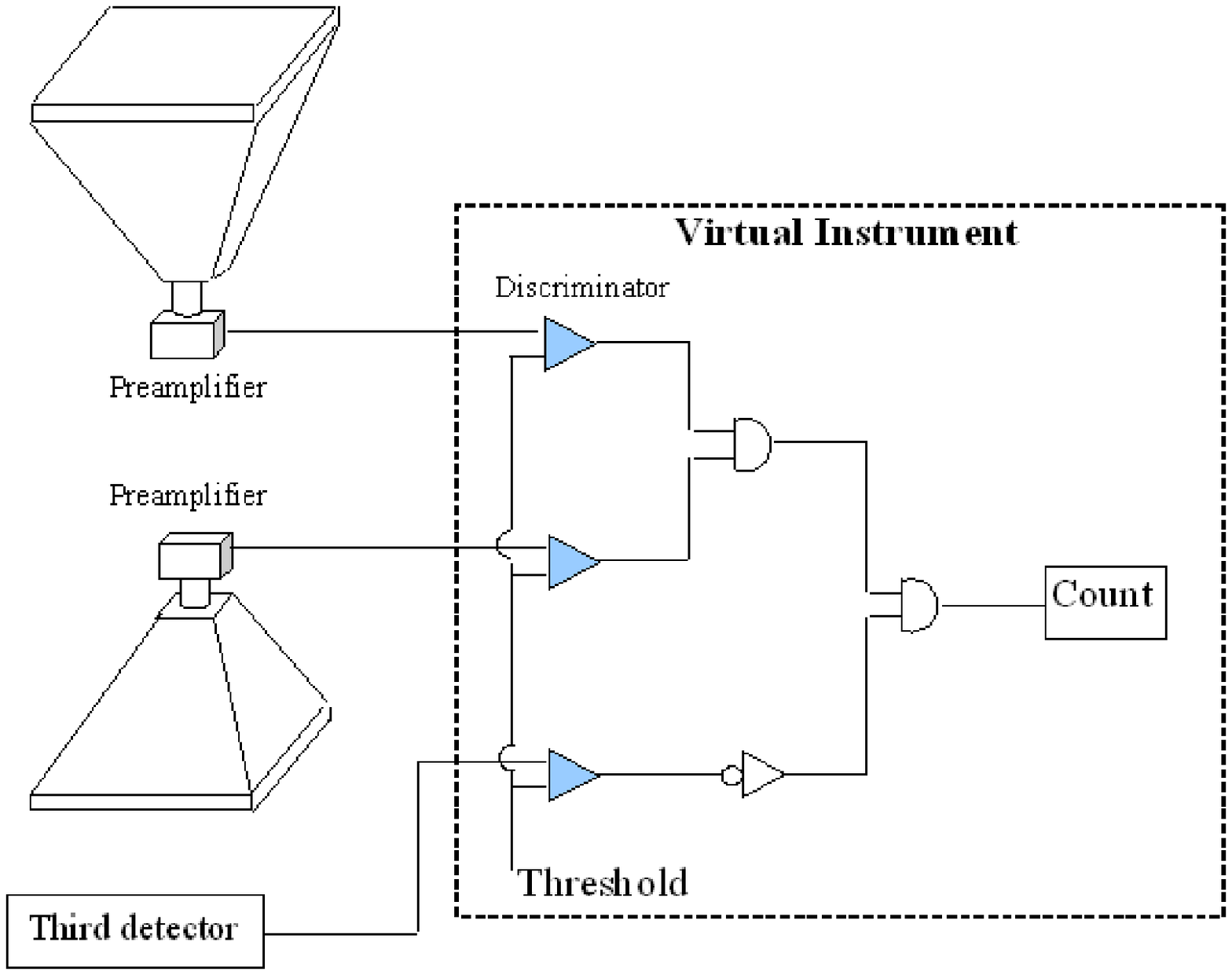}
\vspace*{-7.5cm}
\caption{General layout of the vertical telescope
 including the logic implemented in the data acquisition system
 using LabVIEW software.}
\label{fig1}
\end{figure}  

\begin{figure}[th]
\vspace*{-8.0cm}
\hspace*{-2.0cm}
\includegraphics[clip,width=1.2
\textwidth,height=1.2\textheight,angle=0.] {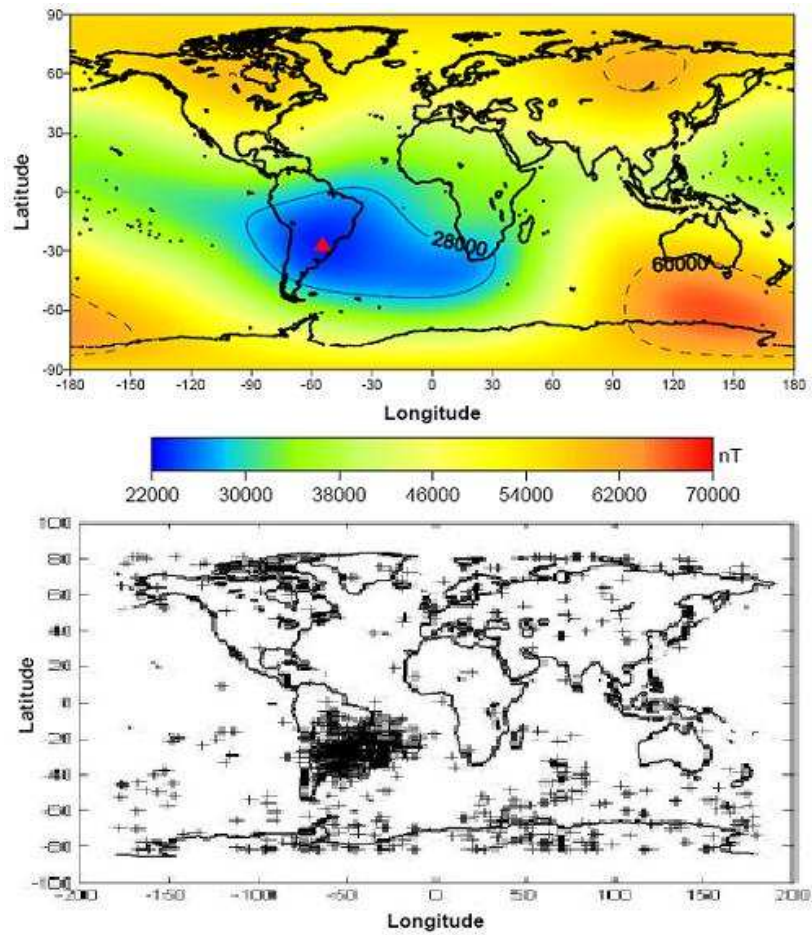}
\vspace*{-6.0cm}
\caption{Top panel: Geographic distribution of geomagnetic field intensity. The SAA boundary is around $B=28000\; nT$. The triangle and the square indicate the localizations of the Tupi facility. Bottom panel: Geographic distribution of single events (particles reaching the Earth) according to MOPPIT space instrument. In the MeV region the background events not connected to the SAA or Polar regions (20\%) are due galactic cosmic rays. Due to particle precipitation, the particle events are 54\% in the SAA and 26\% in the Polar regions. The SAA is the region with the largest event density.}
\label{fig2}
\end{figure} 

\newpage

\begin{figure}
\vspace*{-7.0cm}
\hspace*{+1.0cm}
\centering
\includegraphics[width=\textwidth]{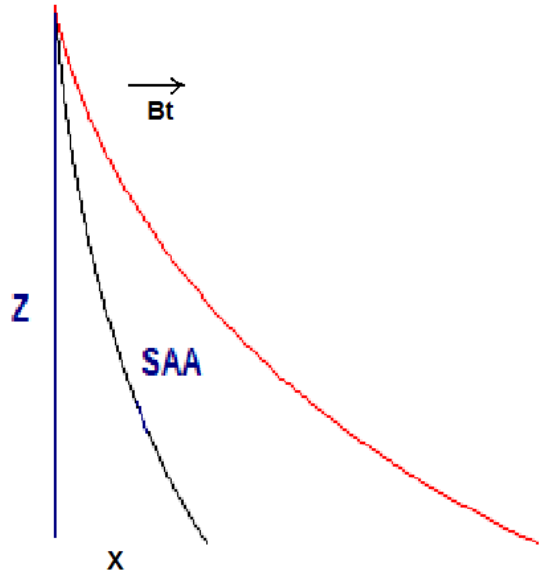}
\vspace*{-6.5cm}
\caption{Comparison between the lateral shift
 of a vertical (born) positive muon
 traveling downward in the atmosphere with momentum $p$,
 due to the Earth's transverse magnetic field, $B_\bot$,
 inside and outside the SAA region.}
\label{fig3}
\end{figure}

\begin{figure}
\vspace*{-8.0cm}
\hspace*{-1.0cm}
\centering
\includegraphics[width=13.9cm]{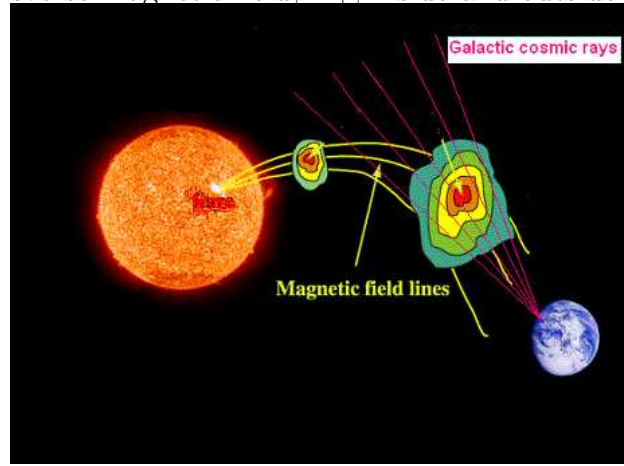}
\vspace*{-6.5cm}
\caption{Diffusive transport between the Sun and the Earth
 of solar charged particles emitted in a solar flare.
 The effect is due to interplanetary magnetic field fluctuations,
 since ions (protons) have a  pitch angle different from zero.
 Consequently, in most cases, only particles produced in very intense flares
 will give some signal (bigger than the background produced by the cosmic rays)
 in the detectors located on the Earth's surface.}
\label{fig4}
\end{figure}   

\newpage

\begin{figure}
\vspace*{-7.0cm}
\hspace*{-1.0cm}
\centering
\includegraphics[width=17.5cm]{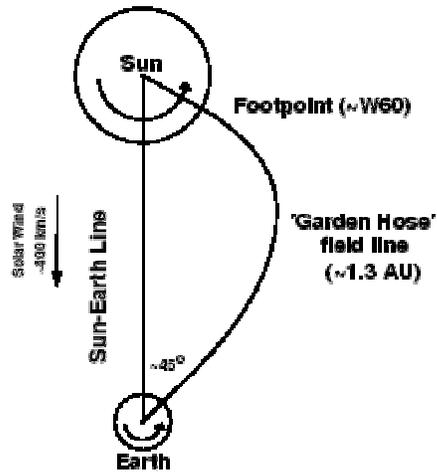}
\vspace*{-15.0cm}
\caption{Example of a good magnetic field connection between the Sun and Earth.
 Flares located near to the foot-point
 of the ``garden hose'' field line between the Sun and Earth
 reach the Earth with a pitch angle close to $45^0$,
 because, protons (ions) travel toward the Earth in a spiral trajectory,
 following the garden hose line and have very sharp onsets.}
\label{fig5}
\end{figure}

\newpage

\begin{figure}
\vspace*{-2.0cm}
\hspace*{-0.0cm}
\centering
\includegraphics[width=\textwidth]{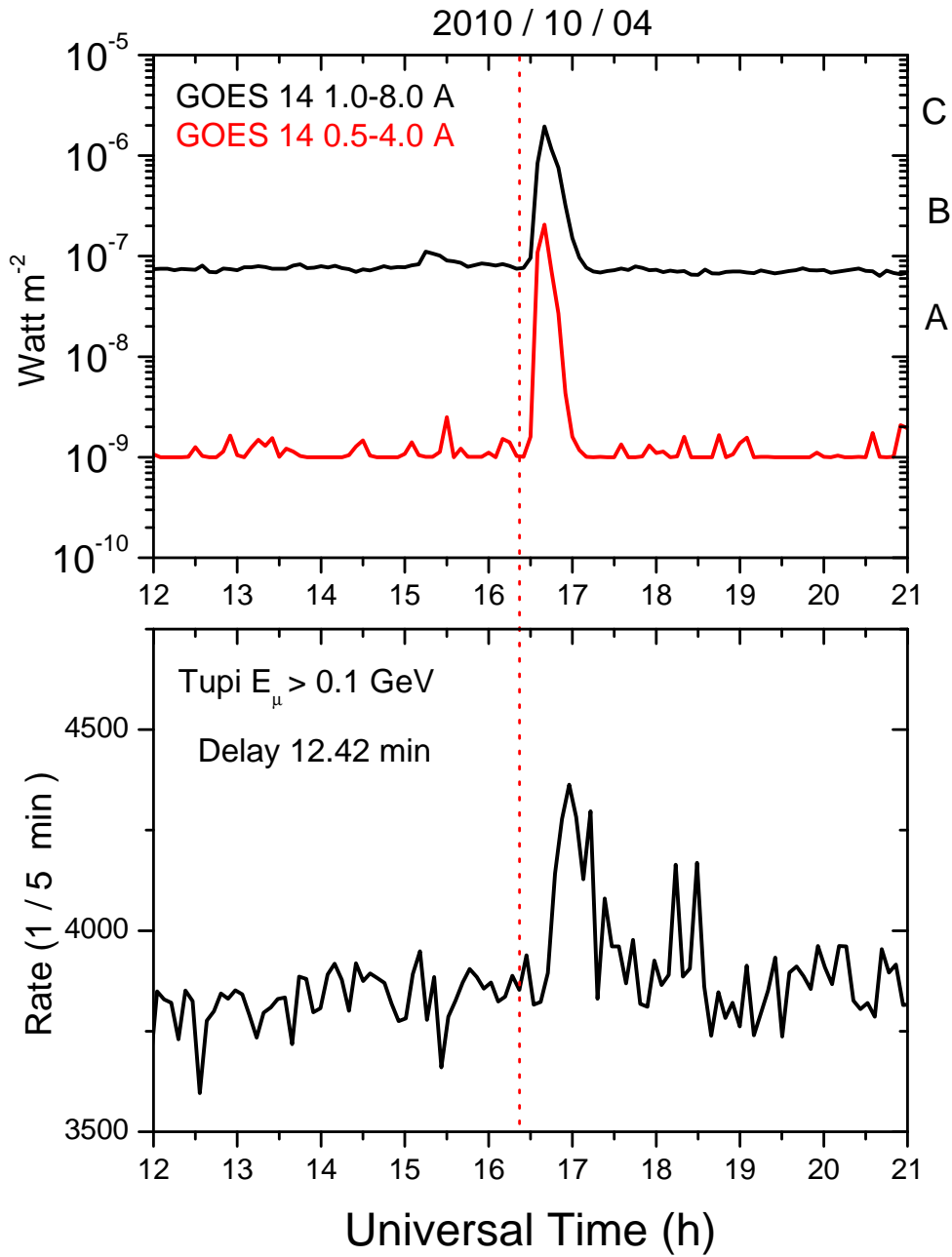}
\vspace*{-2.0cm}
\caption{Top panel: The X-ray flux on July 14, 2010
 according to GOES 14, for two wave lengths.
 Bottom panel: The 5-minute muon counting rate in the vertical Tupi telescope.
 A prompt association around the 21h UT
 between a sharp peak registered by the Tupi telescope
 and a sudden increases in the X-ray emission (mini flare)
 registered by the Goes 14 can be observed.}
\label{fig6}
\end{figure}

\newpage

\begin{figure}
\vspace*{-4.0cm}
\hspace*{-0.0cm}
\centering
\includegraphics[width=\textwidth]{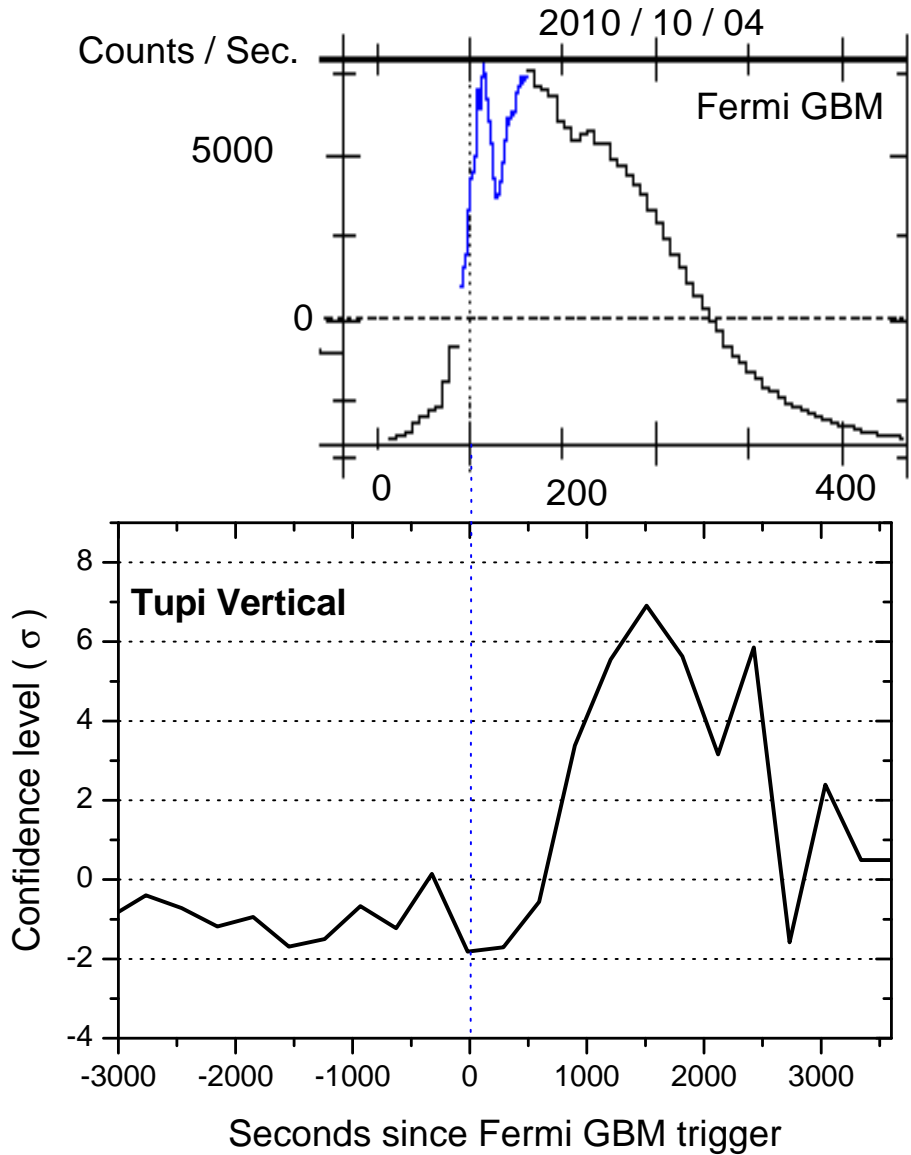}
\vspace*{-2.0cm}
\caption{Top panel: The gamma-ray counting rate on 4 October 2010,
starting at 16:32:31.443 UT  and
 according to Fermi GBM (Trigger bn101004689).
 Bottom panel: The 5-minutes muon counting rate in the vertical Tupi telescope,
 expressed in term of the significance level.}
\label{fig7}
\end{figure}

\newpage

\begin{figure}
\vspace*{-3.0cm}
\hspace*{-0.0cm}
\centering
\includegraphics[width=\textwidth]{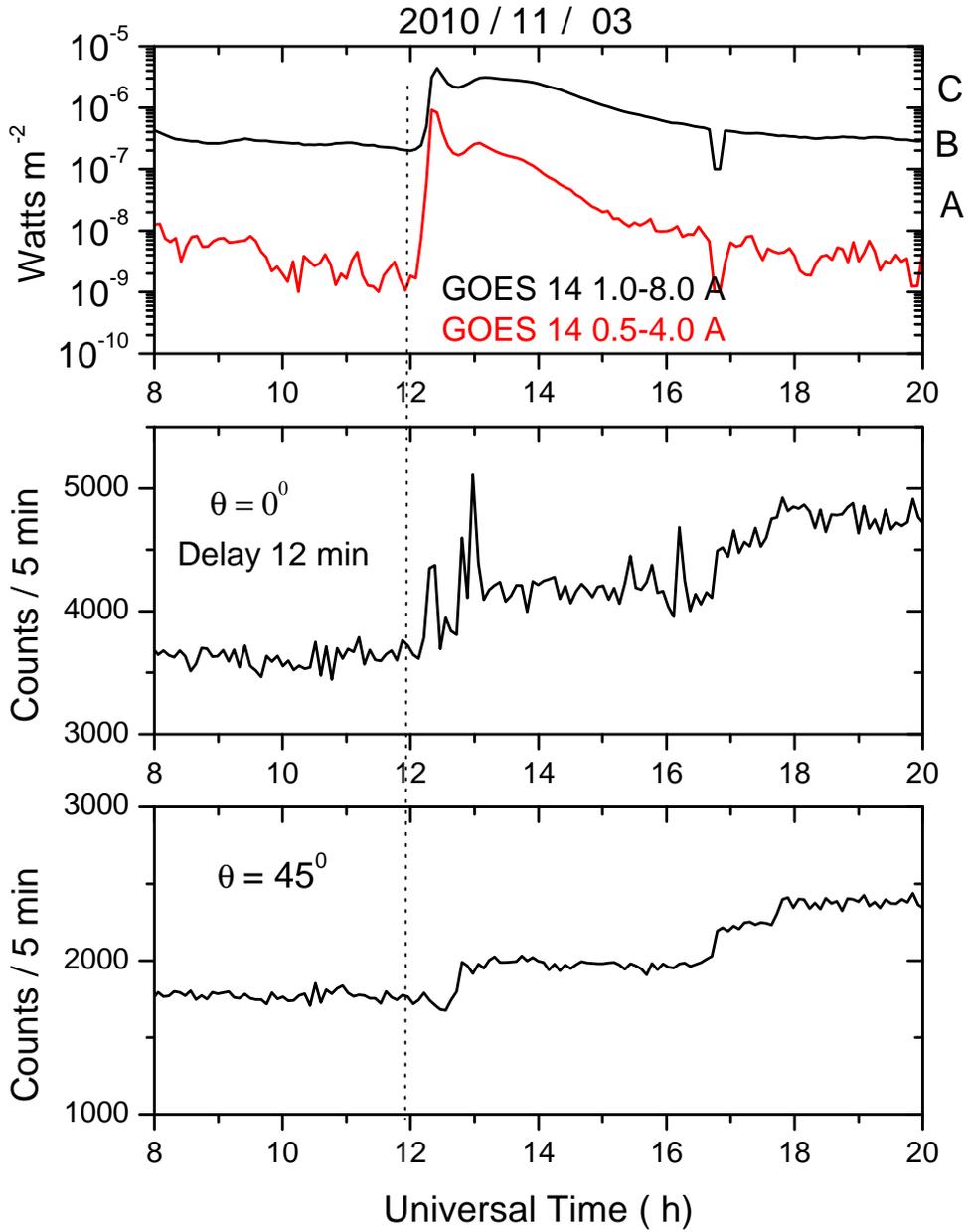}
\vspace*{-2.0cm}
\caption{Top panel: The X-ray flux on 3 November 2010
 according to GOES 14, for two wave lengths.
 Bottom panel: The 5-minute muon counting rate in the vertical Tupi telescope.
 A prompt association around the 21h UT
 between a sharp peak registered by the Tupi telescope
 and a sudden increases in the X-ray emission (mini flare)
 registered by the Goes 14 can be observed, only in the vertical Tupi telescope.}
\label{fig8}
\end{figure}

\newpage

\begin{figure}
\vspace*{-4.0cm}
\hspace*{-0.0cm}
\centering
\includegraphics[width=\textwidth]{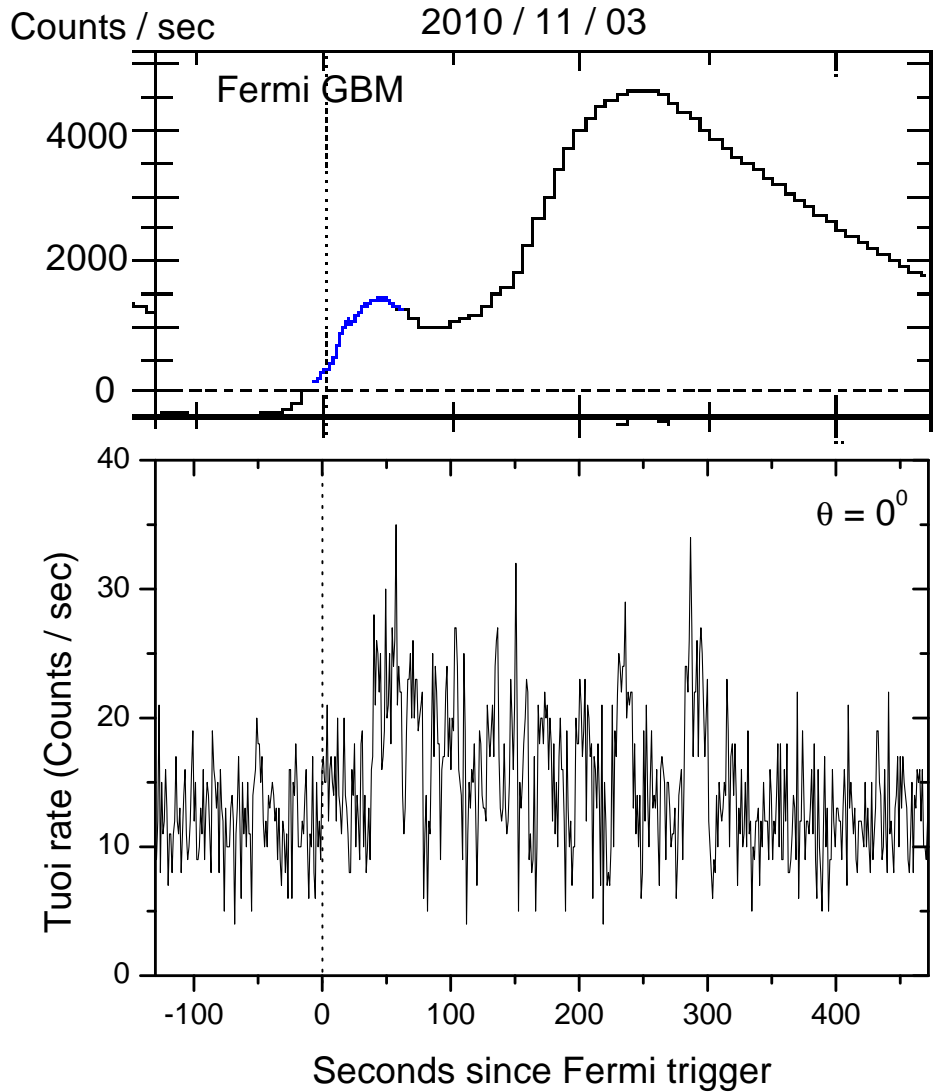}
\vspace*{-1.0cm}
\caption{Top panel: The gamma-ray counting rate on 3 November 2010,
starting at 12:13:19.916 UT  and
 according to Fermi GBM (Trigger bn101103509).
 Bottom panel: The 1-second muon counting rate in the vertical Tupi telescope.}
\label{fig9}
\end{figure}

\newpage

\begin{figure}
\vspace*{-3.0cm}
\hspace*{-0.0cm}
\centering
\includegraphics[width=\textwidth]{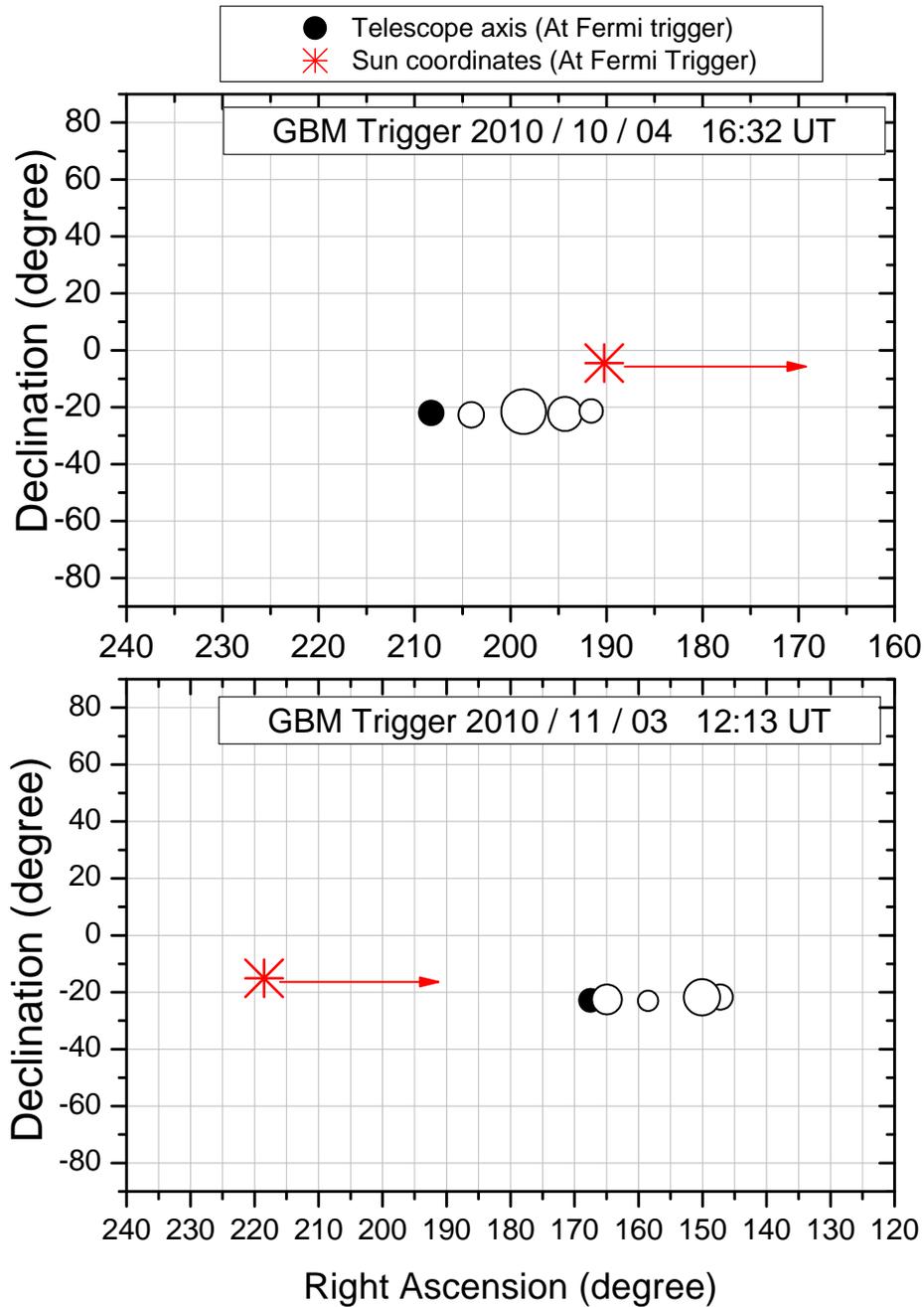}
\vspace*{-2.0cm}
\caption{Sun location, at Fermi trigger time and the muon excess location during the sky scanner by the vertical telescope due to the Earth rotation, for the case of the two flares here analyzed. The size (diameter) of the open circles is proportional to the logarithm of the muon counting rate. In both, the difference between the Sun and telescope axis declination is small, $\Delta \delta = 17.2 $ and  $\Delta \delta = 7.5 $ respectively.
Thus the pitch angle defined as the Sun-ward direction and the telescope axis is approximately equal to the difference between the right ascension of the Sun and of the axis of the telescope.
As one hour in right ascension corresponds to 15 degrees, we observe a pitch angle around 18 degrees for the first event and 51 degrees for the second event.}
\label{fig10}
\end{figure}

\end{document}